\documentclass[aps,floatfix,nofootinbib,preprint]{revtex4}

\usepackage{graphicx}
\usepackage{epic}
\usepackage{eepic}
\usepackage{latexsym}
\usepackage{amssymb,amsmath}

\newcommand{\eq}[1]{(\ref{#1})}
\newcommand{\be}{\begin{equation}}
\newcommand{\ee}{\end{equation}}
\newcommand{\bea}{\begin{eqnarray}}
\newcommand{\eea}{\end{eqnarray}}

\newcommand{\vs}[1]{\vspace{#1 mm}}
\newcommand{\hs}[1]{\hspace{#1 mm}}

\newcommand{\df}{\dot{\phi}}

\def\a{\alpha}
\def\b{\beta}
\def\cc{\gamma}

\def\d{\delta}

\def\e{\epsilon}

\def\f{\phi}
\def\fr{\frac}
\def\F{\Phi}

\def\l{\lambda}
\def\L{\Lambda}

\def\del{\partial}

\let\bm=\bibitem

\begin{document}

\title{Initial Condition Problem is Intractable in Cosmology} 
\author{Ali Kaya}
\email[]{ali.kaya@boun.edu.tr}
\affiliation{Bo\~{g}azi\c{c}i University, Department of Physics, 34342, Bebek, \.{I}stanbul, Turkey \vs{10}}

\begin{abstract}

\vs{5}
	
Determining the initial state of the universe is a challenging problem in quantum cosmology and we argue that the issue is {\it intractable} if the basic postulates of quantum mechanics are {\it not} modified in a nontrivial way. Namely a ``standard" quantum theory of gravity is expected to resolve the big-bang singularity either by yielding a regular past eternal evolution or by a smooth finite beginning; in both cases the initial state can in principle be {\it totally arbitrary}. We illustrate this point in a minisuperspace, gauge fixed, deparametrized toy model where there is a smooth beginning of the universe provided by the matter Hamiltonian degenerating to the zero operator. This arbitrariness is the source of several debates in the literature, especially in relation to inflation, which can only be solved by a new paradigm involving initial conditions that necessarily alters the usual quantum mechanical treatment, but we argue that this is highly improbable.

\end{abstract}

\maketitle

\newpage

\section{A General Discussion}

Quantum mechanics has an incredible success in describing nature over diverse scales from non-relativistic physics to more fundamental field theoretical interactions. Leaving aside technical differences, one essentially applies the same basic rules of quantization in each case. The state of a system is represented as a (unit) vector in a suitable Hilbert space and to each physical observable one associates an operator. The operators corresponding to canonical variables must obey the canonical commutation relations and states evolve according to the Schr\"{o}dinger equation, which requires the specification of an initial wave-function at some time {\it as an input}. The transition amplitude from a given state to another one is given by the Hilbert space inner product. May be the most controversial postulate is related to the measurement process and the collapse of the wave-function, however this can probably be understood without making a distinction between the system and the observer by using ideas like decoherence. There can be alternative (but equivalent) formulations like the path integral approach; or one may demand additional requirements such as relativistic invariance; or there may exist technical complications like the presence of gauge symmetries; in any case the basic postulates are solid and verified with extreme precision by various experiments involving diverse scales. 

It is not clear whether these postulates are also applicable for the quantization of gravity. Perturbation theory on a fixed background geometry fails for general relativity (and also for other theories involving gravity like various supergravities\footnote{Supersymmetric field theories has a nice feature that certain loop infinities cancel out between bosonic and fermionic degrees of freedom running in loops, which is a desirable property for perturbative renormalizability. Unfortunately all supergravities, except may be the maximal $N=8$ one, turn out to be non-renormalizable at some loop order \cite{sg}.}), but this can be viewed as an artifact of a bad approximation. Two leading candidates for quantum gravity are string theory (see e.g. \cite{gsw}) and loop quantum gravity (see e.g. \cite{lqg}). In the later, one indeed directly applies the canonical quantization procedure to general relativity in a non-perturbative and background independent way. Although there are technical peculiarities like introducing a nonseperable (kinematical) Hilbert space,\footnote{This choice radically differs from separable Fock spaces used in quantum field theories, still it can be justifiable since gravity has a distinct role compared to other fields} loop quantum gravity can be viewed as a regular quantum theory as far as the basic postulates are concerned.  

The situation is a bit more involved in string theory. The quantization of free strings propagating in flat space-time is achieved by {\it applying the basic postulates} to the world-sheet string action that yields the spectrum of states. This is true both for closed and open strings, and for D-branes that are described as open strings with Dirichlet boundary conditions. It turns out that the nonlinear Lorentz group generators obey the Lorentz algebra without an anomaly only in certain critical space-time dimensions, which become 26 for bosonic and 10 for supersymmetric strings. At weak coupling, the interactions are introduced by hand using vertex operators, nevertheless the theory eventually yields scattering amplitudes from a {\it given} set of initial to some {\it specified} set of final string states. It is not exactly known how to deal with strings at strong coupling, in that case there are many duality conjectures supported mostly by indirect symmetry arguments. One crucial idea that is embodied as a result of these studies is holography, which states that quantum gravity in $d$ dimensions can equally be described by a $d-1$ dimensional non-gravitational theory \cite{hol1,hol2}. The celebrated AdS/CFT duality \cite{mal} is a concrete example which has important implications. Obviously, non of these astonishing findings indicate a breakdown of the fundamental postulates of quantum mechanics; contrary it is of great importance to substantiate them from basic principles.  

Despite that no change of the axioms of quantum mechanics is foreseen (and apparently needed) for quantization of gravity, it is worth to be cautious in quantum cosmology since the whole universe is now under consideration. As an immediate consequence of this fact, there can be no system/observer (or system/measurement device) division and the postulate involving the collapse of the wave-function is not applicable. Assuming that this issue is understood without the need of introducing different ontologies in the theory, and there are viable suggestions like decoherence as we have pointed out, there is no reason to doubt that the basic postulates of quantum mechanics are not applicable to the whole universe since these are successfully tested for sub-systems with very different scales. 

Of course, there is always the possibility that some postulates of quantum mechanics may need to be modified in the future. No matter what new physics is discovered, it is inevitable that there will always be a state space {\it with potentially infinite content} that is capable of encoding physical properties of systems. Again inevitably, physics will always be about a real or an emergent time evolution in this state space. Hence, the essential question that concerns cosmology is whether one can imagine the existence of a plausible mechanism constraining or completely fixing the initial state of the universe, say at the big-bang. On philosophical grounds, imposing a specific law that is only effective at one time back in the past is questionable. Just the opposite, the laws of physics must be applicable everywhere at all times including the beginning of the universe. Moreover, mathematical consistency comes out as the only criterion in judging the success of such a mechanism which conflicts with one of the basic pillars of science, i.e. testability. Therefore, it is reasonable to think that any presumed law of initial conditions must also be valid today. Yet, such an existing law seems to contradict with all our experience from a wide range of scales, which taught us that initial conditions are free by their nature. 

Without a doubt, the real issue with the big-bang is the presence of a singularity where the curved space-time description breaks down. If general relativity had offered a smooth universe, say cyclically bouncing between two sizes, then there would be no initial condition problem to talk about (instead we would be dealing with philosophical questions like the possibility of {\it actual} past eternal evolution). On the other hand, any new physics in quantum gravity that resolves the big-bang singularity must also work for black-holes or other possible big-crunch/big-rip type cosmological singularities. Hence, it is more likely that the final theory of quantum gravity will offer a smooth description of a gravitating system {\it regardless of its initial conditions.} This would leave just two options for the universe, a regular past eternal evolution or a smooth beginning, and in both cases the initial condition problem becomes intractable.\footnote{A problem in computational complexity theory is called intractable if there exists no efficient algorithm to solve it, which is a statement independent of technological progress.} 

The above discussion does not exclude the possibility that extra and inherently non-dynamical conditions might be imposed on states of gravitating systems for various technical reasons. For example, DeWitt demands the wave-function of the universe to vanish on singular 3-dimensional regions as an attempt to get smooth solutions of the Wheeler-DeWitt (WDW) equation \cite{bdw}. Similarly, there are the no boundary \cite{nb} and the tunneling \cite{t} proposals, which can be viewed as physically viable boundary conditions for the wave-function of the universe (yet, these are recently criticized based on the Lorentzian formulation of quantum cosmology \cite{lor1,lor2}). Nevertheless, neither of these proposals can yield a unique initial state for the universe (one exception that arises in the one-dimensional minisuperspace loop quantum cosmology is discussed in \cite{boj}, but that result does not generalize to the full theory \cite{ash}).

\begin{figure}
\centerline{\includegraphics[width=6cm]{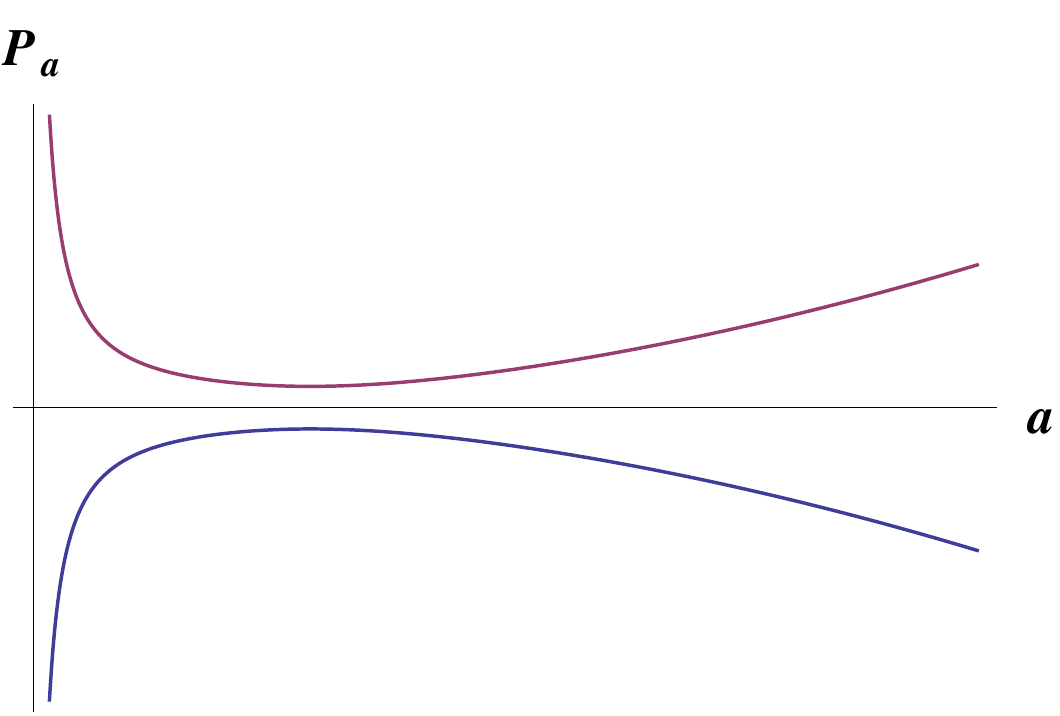}\hs{10}\includegraphics[width=6cm]{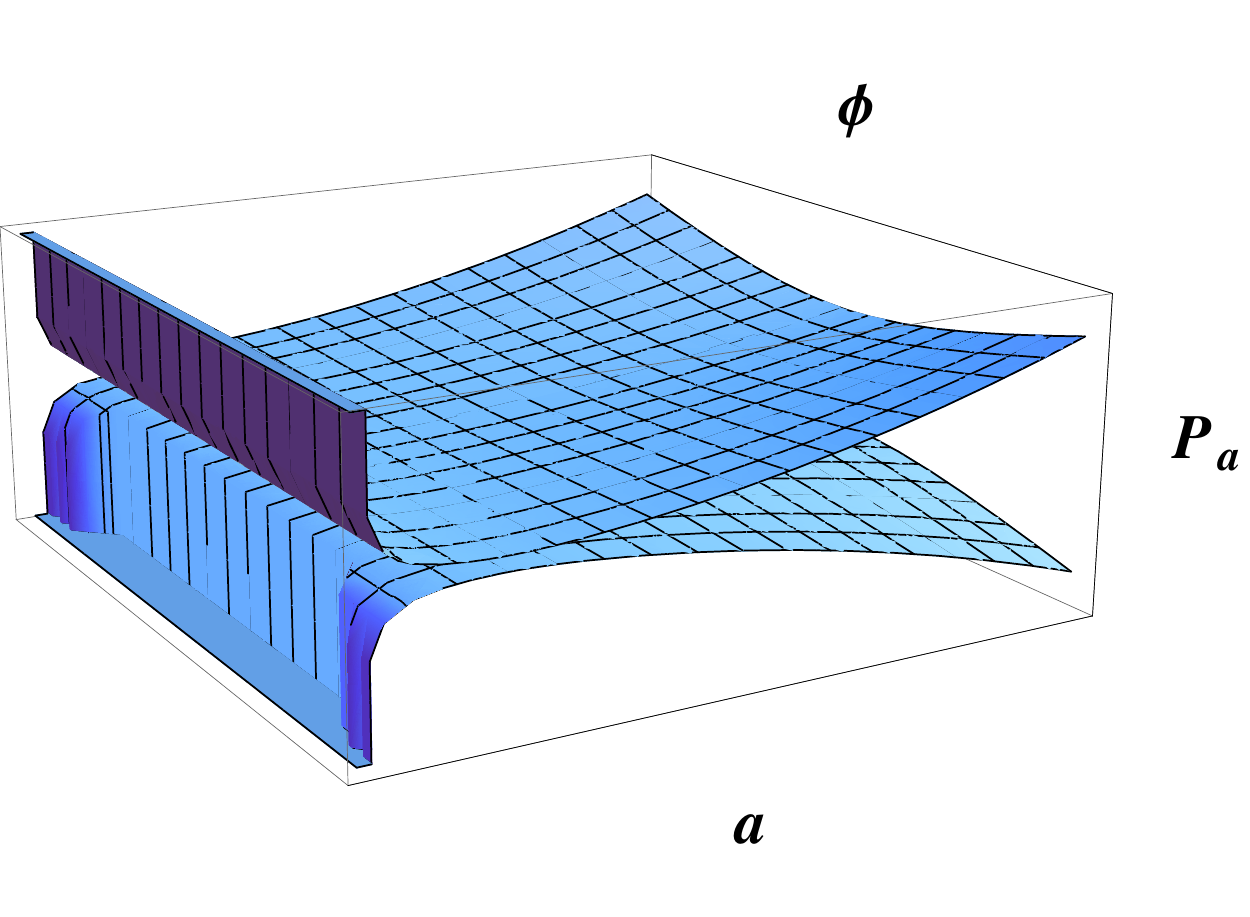}}
\centerline{(a)\hs{70}(b)}
\caption{Sections of the constrained phase space plotted (a) for two fixed values of $\f$ and $P_\f$ (b) for a fixed value of $P_\f$.} 
\label{fig1}
\end{figure}

\section{A Proof of Principle: Minisuperspace Models}

The minisuperspace approximation offers a decent framework for testing ideas in quantum gravity, and it is interesting to study the fate of the big-bang singularity and the associated issue of initial conditions in a simple model where a real scalar field is taken as matter. This problem has been mostly studied using the WDW equation and singularity resolution depends on the implemented boundary conditions, see e.g. \cite{sf}. Although the WDW equation implies interesting physics like the possibility of a signature change (see e.g. \cite{sch1,sch2}), it also faces many issues like the ordering ambiguity in the WDW operator and more importantly the Hilbert space problem that obstructs a clear interpretation of the wave-function (see however \cite{inner} that introduces a set of inner products which may lead to a physically viable probability explanation). 

Nevertheless, there is an alternative and equally legitimate approach where one can deparametrize the theory by imposing a gauge that fixes time and solving the Hamiltonian constraint (different deparametrizations of the scalar field minisuperspace model has been studied in \cite{sf0}). To see how this works, let us start with the minisuperspace action 
\be\label{s1}
S=\int\, dt\, a^3\left[-\fr{6}{N}\fr{\dot{a}^2}{a^2}+\fr{1}{2N}\dot{\f}^2-NV\right],
\ee
where $a(t)$ is the scale factor of the universe, $\f(t)$ is the scalar field, $N$ is the lapse function and $V(\f)$ is the scalar potential (we set the Planck mass $M_p^{-2}\equiv 16\pi G=1$ and absorb the three-dimensional spatial volume of the space in the scale factor so that its mass dimension becomes $[a]=1/M$). The corresponding Hamiltonian can be found as 
\be\label{h}
H=N\left[-\fr{1}{24a}P_a^2+\fr{1}{2a^3}P_\f^2+a^3V(\f)\right]\equiv N\Phi.
\ee
There is a local gauge invariance resulting from purely timelike coordinate transformations and the lapse $N$ is a non-dynamical Lagrange multiplier which imposes the constraint $\F=0$. 

Let us assume that $V>0$, i.e. $V$ is strictly positive. In such a model, the constrained phase space becomes disjoint union of two sub-spaces with $P_a>0$ and $P_a<0$ (see Fig. \ref{fig1}). Crucially $a=0$ should not be included in the phase space since that gives a singular metric. The fact that $P_a$ never vanishes allows one to use the conjugate variable $a$ as the time coordinate as a gauge choice. Solving $P_a$ from $\F=0$ (with the negative root giving $\dot{a}>0$) and using $a$ as time gives the following reduced action in the Hamiltonian form \cite{a1}
\be\label{s2}
S=\int da\left[P_\f \fr{d\f}{da}-\fr{\sqrt{12}}{a}\sqrt{P_\f^2+2a^6V(\f)}\right].
\ee
Carrying out above deparametrization in the phase space path integral quantization, one finds that the Faddeev-Popov determinant corresponding to the gauge fixing that uses $a$ as the time coordinate exactly cancels out the term that arises after $P_a$ integration, which is actually constrained by the delta functional $\d(\F)$ \cite{a1}. Therefore, \eq{s2} is also {\it exact} in the quantum theory. 

It is important to emphasize that \eq{s2} governs the dynamics in the expansion branch of the phase space and there is a corresponding action for the contracting branch. In the quantum theory, these two branches are {\it completely split up} since there is no (on-shell or off-shell) classical configuration connecting them, as a result, no tunneling is possible. Hence, the situation is similar to a system of two separated infinite square-wells where no tunneling is possible once a particle is placed in one of the wells.   

After deparametrization, one ends up with a one dimensional system with the phase space coordinates $(\f,P_\f)$ and the following time-dependent Hamiltonian (note that $a$ plays the role of time here) 
\be\label{h2}
\mathcal{H}=\fr{\sqrt{12}}{a}\sqrt{P_\f^2+2a^6V(\f)}.
\ee
The big-bang occurs at $a=0$ and not surprisingly the classical dynamics becomes singular at that moment. For quantization, one can introduce the Hilbert space of square integrable functions $\psi(\f)$ over the variable $\f\in(-\infty,+\infty)$ and represent the momentum operator as usual by $P_\f=-i\del_\f$. If $V\to\infty$ as $\f\to\pm\infty$ then at any given $a>0$ the operator $P_\f^2+2a^6V(\f)$ becomes self-adjoint with a discrete positive definite point spectrum. As a result the Hamiltonian \eq{h2} becomes uniquely defined by the square root of this operator, which is completely similar to determining the square-root of a Hermitian matrix. The fate of the singularity in quantum theory now depends on the behavior of $\mathcal{H}$ as $a\to0$. 

For an exactly solvable example one may take
\be\label{pot}
V(\f)=\fr12 m^2\f^2+\L,
\ee
where $\L>0$ is a (cosmological) constant introduced to satisfy $V>0$. In that case, $\mathcal{H}$ becomes the square root of the standard harmonic oscillator Hamiltonian whose spectrum can readily be calculated as 
\be\label{spect}
E_n=\fr{\sqrt{24}}{a}\sqrt{ma^3\left(n+\fr12\right)+a^6\L},\hs{5}n=0,1,2,3...
\ee
so that $\lim_{a\to0}E_n\to0$. The corresponding eigenstates are 
\be\label{ef}
\psi_n=\fr{\sqrt{\a}}{\sqrt{2^nn!\sqrt{\pi}}}\,H_n(\a\f)\,e^{-\fr12\a^2\f^2},\hs{3}n=0,1,2...
\ee
where $\a=\sqrt{ma^3}$ and $H_n$ are Hermite polynomials. Thus, $\mathcal{H}$ indeed degenerates to the zero operator at the big-bang without any singular behavior. 

Technically, one should a bit be careful about the last statement since the eigenstates \eq{ef} also degenerate as $a\to0$. Yet, by expressing the operator \eq{h2} in a regular basis $\left.|\l_n\right>$, one may find that the matrix entries satisfy $\lim_{a\to0}\left<\l_n|\mathcal{H}|\l_m\right>\to0$ \cite{a1}. Indeed, choosing for instance 
\be\label{el}
\l_n=\fr{1}{\sqrt{2^nn!\sqrt{\pi}}}\,H_n(\f)\,e^{-\fr12\f^2},\hs{3}n=0,1,2...
\ee
which form a complete orthonormal set of basis vectors in the Hilbert space satisfying $\left<\l_n|\l_m\right>=\d_{nm}$ and $I=\sum_n\left|\l_n\right>\left<\l_n\right|$, the matrix entries of $\mathcal{H}$ can be calculated as
\be\label{28}
\mathcal{H}_{nm}\equiv\left<\l_n|\mathcal{H}|\l_m\right>=\sum_{k=0}^\infty\left<\l_n|\psi_k\right>\left<\psi_k|\l_m\right>E_k.
\ee
Using the functions \eq{ef}, \eq{el} and the spectrum $E_k$ given in \eq{spect} one finds (after introducing $M_p$ factors) that 
\be\label{beha0}
\lim_{a\to0}\mathcal{H}_{nm}=(mM_p^2)\,a^2\left[C_{nm}+{\cal O}(a)\right],
\ee
where $C_{nm}$ are $a$ independent. Consequently, one sees that $\mathcal{H}$ really degenerates to the zero operator as $a\to0$.

It is possible to rigorously study the evolution of the state vectors in the limit $a\to0$ by referring to the above completely regular basis \eq{el} in the Hilbert space. Using the unitary evolution operator 
\be\label{u}
U(a_2,a_1)=T\exp\left[-i\int_{a_1}^{a_2}\mathcal{H}(a)da\right]
\ee
where $T$ denotes time ordering, the Schr\"{o}dinger equation 
\be\label{sce}
i\del_a\psi=\mathcal{H}\psi
\ee
can be solved as $\psi(a_2)=U(a_2,a_1)\psi(a_1)$. Expressing $U(a_2,a_1)$ in the basis \eq{el} and using the behavior found in \eq{beha0}, one can see that $a_1\to0$ limit is perfectly well defined in \eq{u}. Thus, for {\it any given initial state} $\psi(0)$ at $a=0$ in the Hilbert space, the Schr\"{o}dinger equation $i\del_a\psi=\mathcal{H}\psi$ can be solved {\it uniquely} as 
\be
\psi(a)=U(a,0)\psi(0),
\ee
which determines the wave-function of the universe $\psi(a)$ at a later time. As a result there is no technical or physical reason for the initial wave-function not to be an arbitrary state \cite{a1}. 

Curiously, this model can be viewed as a toy realization of the zero energy hypothesis, which assumes that the universe was born as a zero energy quantum fluctuation \cite{tr}. It will be interesting to study perturbations around the homogeneous background to see whether their Hamiltonian also degenerates to zero as $a\to0$. If this turns out ot be the case, one would get a much more realistic realization of the zero energy proposal. As shown in \cite{lor2}, the primordial cosmological perturbations may be unsuppressed in the Lorentzian formulation of quantum cosmology yielding a severe backreaction problem in some models. Naturally, one would expect no serious backreaction issue to arise if the Hamiltonian of the perturbations vanishes as $a\to0$.  

This simple example nicely illustrates how quantum mechanics can resolve the big-bang singularity and how this makes the initial condition problem intractable. The only remaining option for progress is to make an educated {\it assumption} for the initial state of the universe, which is indeed the main practice in quantum cosmology. 

One may concern that the above results might be specific to a simple model. Yet, it is not difficult to see that similar conclusions also hold in the most general spatially flat minisuperspace model \cite{a1}. For that case the metric can be written as  
\be
ds^2=-N^2dt^2+h_{ij}(t)\,dx^idx^j
\ee
and it is convenient to introduce the following decomposition 
\be
h_{ij}=a^2\,\cc_{ij},
\ee
where $\det \cc_{ij}=1$ and $\det h_{ij}=a^6$. A straightforward calculation gives the following minisuperspace action
\be\label{sn2}
S=\int\, dt\, a^3\left[-\fr{6}{N}\fr{\dot{a}^2}{a^2}+\fr{1}{2N}g_{\a\b}\dot{u}^\a\dot{u}^\b+\fr{1}{2N}\dot{\f}^2-NV\right],
\ee
where $u^\a$ ($\a,\b=1,..,5$) denote appropriate variables that parametrize general unit determinant matrices\footnote{For instance, one may write $\cc_{ij}=\exp\left[u^\a t_\a\right]_{ij}$ where $(t_\a)_{ij}$ is a chosen basis in the space of trace free matrices.} $\cc_{ij}=\cc_{ij}(u)$  and $g_{\a\b}(u)=\fr12 (\del_\a\cc_{ij})(\del_\b\cc_{kl})\cc^{ik}\cc^{jl}$ is a {\it positive definite} metric in the $u^\a$-parameter space. The corresponding Hamiltonian can be found as 
\be\label{h3}
H=N\left[-\fr{1}{24a}P_a^2+\fr{1}{2a^3}g^{\a\b}P_\a P_\b+\fr{1}{2a^3}P_\f^2+a^3V(\f)\right],
\ee
where $P_\a$ is the momentum conjugate to $u^\a$ and these obey the Poisson bracket  $\{u^\a,P_\b\}=\d^\a_\b$. 

For $V>0$, the phase space again decomposes into two disjoint sub-spaces. After setting $a=t$ as a gauge condition and solving $P_a$ from \eq{h3}, one may obtain the reduced Hamiltonian in the expanding branch of the reduced phase space as  
\be\label{h4}
\mathcal{H}=\fr{\sqrt{12}}{a}\sqrt{g^{\a\b}P_\a P_\b+P_\f^2+2a^6V(\f)}.
\ee
Here, the matter $(\f,P_\f)$ and the gravitational $(u^\a,P_\a)$ degrees of freedom decouple from each other and the full Hilbert space takes the form of a tensor product of the two Hilbert spaces corresponding to $\f$ and $u^\a$ variables. Unfortunately, there is now an ordering ambiguity in the gravitational sector and traditionally this is bypassed by choosing an ordering that makes $g^{\a\b}P_\a P_\b$ self-adjoint in the quantum theory. Assuming that this operator has a discrete spectrum with ($a$-independent) eigenstates $\psi_r$ and eigenvalues $\e_r$, one may introduce the following orthonormal basis 
\be\label{tb}
\Psi(u,\f)_{(r,n)}=\psi_r(u)\otimes \tilde{\psi}_n(\f)
\ee
in which the Hamiltonian \eq{h4} becomes diagonal. The (instantaneous) eigenvalues of $\mathcal{H}$ for the potential \eq{pot} can be found as
\be
E_{(r,n)}=\fr{\sqrt{12}}{a}\sqrt{\e_r+2m a^3\left(n+\fr12\right)+2a^6\L},
\ee
which can be compared to \eq{spect}.

The time evolution simplifies near the big-bang singularity since the matter Hamiltonian degenerates to the zero operator. In that case the Schr\"{o}dinger equation $i\del_a\Psi=\mathcal{H}\Psi$ can be solved approximately to yield
\be
\lim_{a\to0}\Psi(u,\f,a)\simeq \sum_r c_r \exp\left[-i\sqrt{12\e_r}\ln a\right]\psi_r\otimes \tilde{\psi}(\f),
\ee
where $c_r$ and $\tilde{\psi}(\f)$ are completely arbitrary. Therefore the singularity at $a=0$ gives diverging phases, which is typical in cosmology as discussed in \cite{a2}. If the evolution starts with a single energy eigenstate at $a=0$ where all but one of the $c_r$ coefficients are zero, the infinite phase becomes overall and harmless since it cancels out in the expectation values.\footnote{It is not possible to cure the infinite phase problem when the spectrum of $g^{\a\b}P_\a P_\b$ is purely continuous since then any proper state in the Hilbert space necessarily becomes a superposition of the improper eigenvectors.} This is an example where the resolution of the big-bang singularity restricts the form of the initial wave-function as envisaged by DeWitt \cite{bdw}, nevertheless there is still a huge arbitrariness related to the choice of the nonzero $c_r$ coefficient and the matter state $\tilde{\psi}(\f)$, which are left over. 

\section{Discussion}

\begin{figure}
\centerline{\includegraphics[width=8cm]{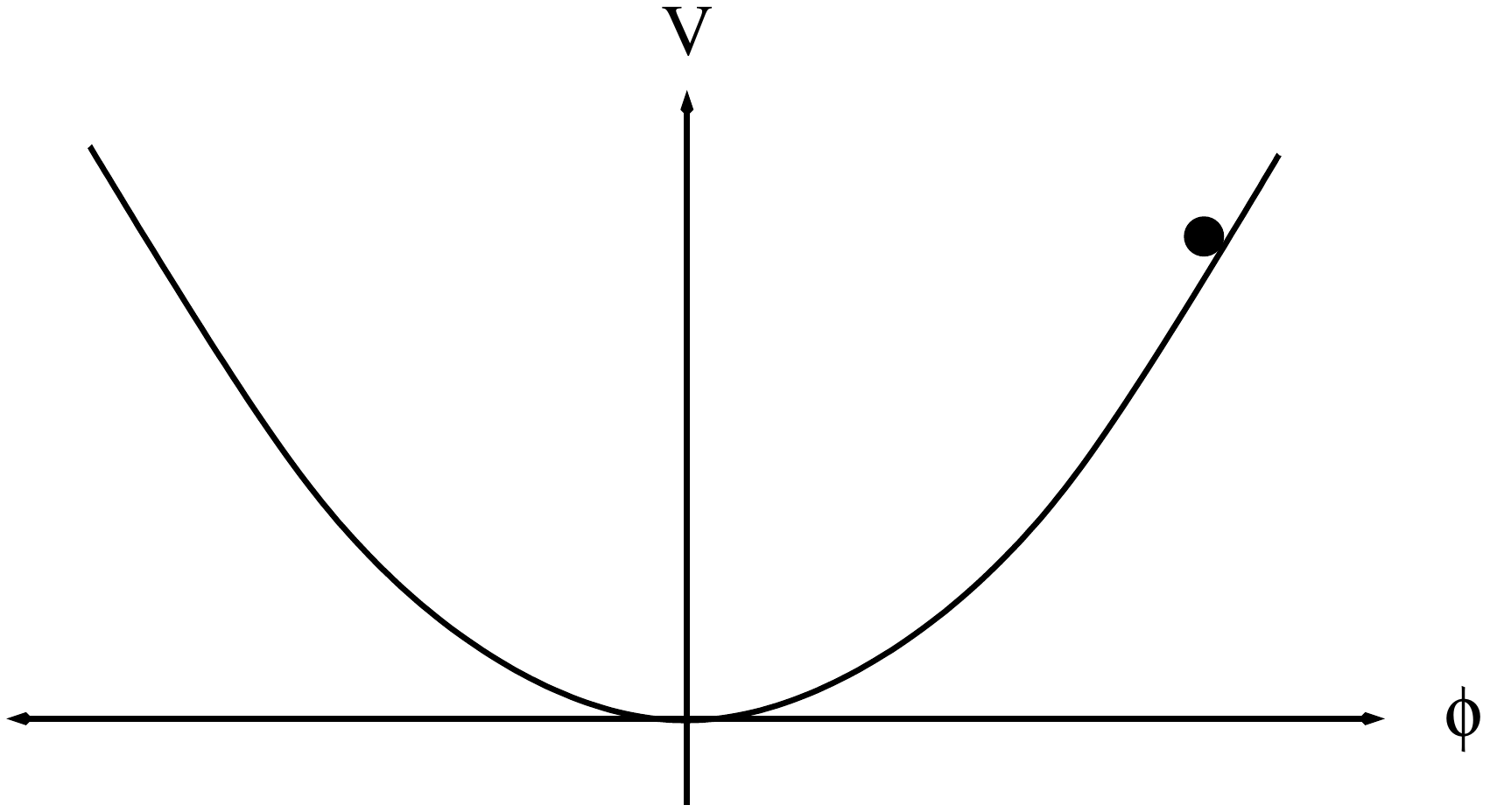}}
\caption{A scalar field driven chaotic inflationary model with the potential $V=\fr12 m^2\f^2$. For slow-roll inflation to occur, the scalar must be released at large values $\f\gg M_p$ with negligible speed $\df$, where $M_p$ is the Planck mass.} 
\label{fig2}
\end{figure}

It is interesting to examine how the probability of inflation can be determined in the above simple but exactly solvable model, where the potential \eq{pot} actually gives slow-roll chaotic inflation if the inflaton is {\it semi-classically localized} around a large value $|\f|\gg M_p$, see Fig \ref{fig2} (we assume that $\L$ is negligibly small in this discussion). As the most reasonable try, one may identify the probability of inflation as $p=1-\int_{-\a M_p}^{\a M_p}\,|\psi(\f)|^2\,d\f$ where $\psi(\f)$  is a given normalized state and $\a\gg1$ is a constant so that $\f\geq\a M_p$ ensures the onset of inflation. This however contradicts with the requirement that the scalar must be semi-classically localized around a definite value since $\f$ is actually {\it fluctuating} in a generic state $\psi(\f)$ and the corresponding gravitational back-reaction is not the same with a scalar field having a definite value \cite{mwu}. One may then think that a well peaked Gaussian state centered around some $|\f|\gg M_p$ can definitely yield inflation. Still, the question why the initial wave-function of the universe should be such a Gaussian remains unanswered since there turns out to be no constraints for the initial state as we have discussed. 

As an attempt to solve these unavoidable difficulties, one then supposes the existence of a multiverse which offers a large (or even an infinite) number of baby universes to {\it play with.} One imagines that in the multiverse the scalar field takes random values (or states) in different baby universes so that inflation inevitably occurs in some of them (hence the name of the model i.e. chaotic inflation). However, the multiverse idea creates more questions than it answers. What determines the state in each baby universe still begs for an answer, i.e. each baby universe now has its own initial condition problem. It is not clear why the states are randomly assigned or why they are distributed according to a probability distribution. Why baby universes are treated like independent members of an ensemble as if they are identical copies of a statistical mechanical system in equilibrium? Moreover, there are infinite possibilities for initial conditions and thus even for an infinite multiverse it is difficult to develop a meaningful theory due to the inherent mathematical ambiguities involving infinities. 

Based on the arguments similar to the ones elaborated above, the theory of inflation has been severely criticized in \cite{cc1} by claiming that it has no predictive power. This criticism has been refuted in an unconventional way by a group of 33 prominent physicists \cite{cc2}. We think that our point of view presented in this paper explains the reason for the disagreement. Inflation is definitely predictive as long as one assumes suitable initial conditions giving an almost exponential expansion and furthermore supposes that quantum fields on this background are released in their Bunch-Davies vacua.\footnote{This second assumption, which is also essential for inflation to produce scale free cosmological perturbations and hence to be consistent with observations, is mostly overlooked in the discussions about the plausibility of inflation. However, there is no a priori reason for quantum fields to be released in their Bunch-Davies vacua since in a time dependent background there is no unique preferred vacuum state, on the contrary, one can only define an instantaneous vacuum for each time. Indeed, even in the presence of a preferred vacuum, starting the system from that state is a pure choice.} There is no way to justify these assumptions from first principles but this is not a valid point of criticism since any theory of cosmology will certainly suffer from similar shortcomings about initial conditions.

\end{document}